\documentclass[a4paper,11pt]{article}
\usepackage{pos}

\title{Prospects for the stout smearing as an equivalent approach to the Wilson flow}

\author*[a]{Masato Nagatsuka}
\author[a]{Keita Sakai}
\author[a]{Shoichi Sasaki}

\affiliation[a]{Department of Physics, Tohoku University, Sendai 980-8578, Japan}

\emailAdd{masaton@nucl.phys.tohoku.ac.jp}
\emailAdd{sakai@nucl.phys.tohoku.ac.jp}
\emailAdd{ssasaki@nucl.phys.tohoku.ac.jp}

\abstract{We present the equivalence between the Wilson flow and the stout smearing. The similarity between these two methods was first pointed out by L\"uscher’s original paper on the Wilson flow. We first show the analytical equivalence of two methods, which indicates that the finite stout smearing parameter induces ${\cal O}(a^2)$ correction. We secondly show that they remain equivalent in numerical simulations within some numerical precision even with finite cutoffs and stout smearing parameters by directly comparing the expectation values of the action density and we shortly mention the use of the equivalence.}

\FullConference{The 40th International Symposium on Lattice Field Theory (Lattice 2023)\\
July 31st - August 4th, 2023\\
Fermi National Accelerator Laboratory\\}

\begin{document}
\maketitle

\section{Introduction}
\label{sec:INTRO}
The Yang-Mills gradient flow ~\cite{{Narayanan:2006rf},{Luscher:2009eq},{Luscher:2010iy}} is described by equations that diffuse the gauge field smoothly as a function of fictitious time denoted as flow time. Since the Yang-Mills gradient flow for a positive flow time provides us the ultraviolet finite correlation functions made of the flowed gauge field to all orders in perturbation theory without any multiplicative wave function renormalization~\cite{Luscher:2011bx},
it is extensively used in, beyond its proposal, e.g. high-precision reference scale determination~\cite{{Borsanyi:2012zs},{Sommer:2014mea}}, computing the nonperturbative running of the coupling constant~\cite{Ramos:2014kla} and chiral condensate~\cite{Luscher:2013cpa}, defining the energy-momentum tensor on the lattice~\cite{Suzuki:2013gza}, and so on~\cite{Luscher:2013vga}. 

The Yang-Mills gradient flow on the lattice field theory is called the Wilson flow, which is described by the following differential equation with the initial conditions $V_\mu(0,x)=U_\mu(x)$,
%
%
\begin{equation}
\frac{\partial}{\partial t}V_\mu(t,x)=-g_0^2 \partial_{x,\mu}S_{W}[V] V_\mu(t,x).
\label{eq:Wilson_flow}
\end{equation}

On the other hand, the stout smearing is a smoothing technique of gauge field and has the following analytical and iterative expression:

\begin{equation}
U_{\mu}^{(k+1)}(x)=\exp\left(
i\rho Q^{(k)}_{\mu}(x)
\right)U_{\mu}^{(k)}(x),
\label{eq:new_link_stout}
\end{equation}
which means the link variables $U_{\mu}^{(k)}(x)$ at step $k$ are mapped into the link variables $U_{\mu}^{(k+1)}(x)$.
The $Q^{(k)}_{\mu}(x)$ corresponds to a Lie algebra valued quantity given by 
%
%
\begin{equation}
Q^{(k)}_{\mu}(x)=ig_0^2 \partial_{x,\mu}S_{W}[U_{\mu}^{(k)}(x)]
\label{eq:lie_algebra_value}
\end{equation}
with the Wilson action in terms of the stout link $U_{\mu}^{(k)}(x)$. 
L\"uscher pointed out that
the Wilson flow can be regarded as a continuous version of
the stout smearing~\cite{{Luscher:2010iy},{Luscher:2009eq}}. 
Although, this idea was followed in computing topological observables~\cite{{Bonati:2014tqa},{Alexandrou:2017hqw}},
the classical $a$-expansion is used.

Therefore, we examine the equivalence of
the two methods from two aspects: First, we check the rigorous analytical equivalence at a proper limit. Second, we measure the expectation value of the action density $\langle E\rangle$ to verify whether they remain equivalent in numerical simulations. 

\section{Outline of technical details}
\label{sec:FLOW_STOUT}

Recently, two of our collaborators found the numerical equivalence between the spatial Wilson flow~\footnote{The diffusion is restricted only to spatial directions.} and the stout-link smearing in the glueball spectroscopy~\cite{{Sakai:2022zdc},{Sakai:2022mgd}}. 
To understand the numerical results, we explicitly derive a continuous version of the stout smearing procedure {\it at finite lattice spacing $a$} in the limit of $\rho\to0$ that leads to the Wilson flow as an extension of Refs.~\cite{{Luscher:2009eq},{Bonati:2014tqa}}.
In this section, we briefly see
technical details, which are discussed in Ref.~\cite{{Sakai:2022zdc},{Nagatsuka:2023}}.

First, we subtract $U^{(k)}_\mu(x)$ from both sides of Eq.~\eqref{eq:new_link_stout},

%
%
\begin{align}
\frac{U^{(k+1)}_\mu(x)-U^{(k)}_\mu(x)}{\rho}=&\frac{1}{\rho}\left(\exp\left(i\rho Q_{\mu}^{(k)}(x)\right)-1\right)U^{(k)}_\mu(x) \nonumber \\
=&iQ_\mu^{(k)}
\left(
\sum_{n=0}^{\infty}\frac{(i\rho Q_\mu^{(k)})^n}{(n+1)!}
\right)U^{(k)}_\mu(x).
\end{align}
taking the limit of $\rho\to0$ leads to the following differential equation
%
%
\begin{equation}
  \label{eq:stout_flow_equation}
  \frac{\partial}{\partial s}\tilde{U}_\mu(s, x)=iQ_\mu(s,x)\tilde{U}_\mu(s, x),
\end{equation}
where we replaced $U^{(k)}_\mu(x)$ with $\tilde{U}_\mu(s, x)$ and $Q^{(k)}_\mu(x)$ with $Q_\mu(s, x)$ according to the conversion ${s=\rho k}$. 
Second, using the explicit expression for $Q^{(k)}_{\mu}(x)$ and $S_{W}[U_{\mu}^{(k)}(x)]$, we derive Eq.~\eqref{eq:lie_algebra_value} which was originally derived by L\"uscher in Ref.~\cite{Luscher:2009eq} and its derivation was sophisticated in Ref.~\cite{Bonati:2014tqa}.
Finally, by combining with Eq.~\eqref{eq:lie_algebra_value} and Eq.~\eqref{eq:stout_flow_equation}, we get
%
%
\begin{equation}
\label{eq:stout_differential_eq}
\frac{\partial}{\partial s}\tilde{U}_\mu(s,x)=-g^2_0\partial_{x,\mu}S_{W}[\tilde{U}]\tilde{U}_\mu(s,x),
\end{equation}
which is exactly same as the Wilson flow equation~\eqref{eq:Wilson_flow} 
under the correspondences of $t \leftrightarrow s$ and $V_{\mu}(t,x) \leftrightarrow \tilde{U}_{\mu}(s,x)$.
Remark that we can rigorously identify $\rho n_{\rm st}$, where $n_{\rm st}$ denotes smearing steps, as the flow time $t$ at finite lattice spacing $a$ which was supposed to be the perturbative matching relation in Refs.~\cite{{Thomas:2014tda},{Alexandrou:2017hqw}}.

It is worth emphasizing that Eq.~\eqref{eq:stout_differential_eq} is derived in the limit of $\rho\to 0$ without classical $a$-expansion and that, for the finite smearing parameter $\rho$, the leading order corrections on Eq.~\eqref{eq:stout_differential_eq} 
start at ${\cal O}(\rho)$, which induces ${\cal O}(a^2)$ corrections since the flow time $t=\rho n_{\rm st}$ has dimension length squared~\cite{Luscher:2010iy}.
Thus, we can numerically reproduce the Wilson flow 
from the stout smearing even at finite lattice spacing $a$ as long as we choose the enough small smearing parameter $\rho$.


\section{Numerical results} 
\label{sec:Num_results}
%
%
\begin{table*}[b]                                                    
  \caption{
  Summary of the gauge ensembles: gauge coupling, lattice size ($L^3\times T$), plaquette value, 
  lattice spacing ($a$), spatial extent ($La$), the Sommer scale ($r_0$), 
  the number of the gauge field configurations ($N_{\rm conf}$),  
  the number of thermalization sweeps ($n_{\rm therm}$)  
  and the number of update sweeps ($n_{\rm update}$).
  All lattice spacings are set by the Sommer scale ($r_0=0.5$ fm)~\cite{{Sommer:1993ce},{Necco:2001xg}}.
  \label{tab:Sims}
  }
\small
\begin{tabular}{ c c c c c c c c c }
\hline
$\beta=6/g_0^2$ & $L^3\times T$ & plaquette & $a$ [fm] & $La$ [fm] & $r_0/a$ (Ref.~\cite{Necco:2001xg}) & $N_{\rm conf}$ & $n_{\rm therm}$ & $n_{\rm update}$ \cr 
\hline
5.76 & $16^3\times 32$ & 0.560938(9)     & 0.1486(7)    & 2.38   & 3.364(17)     & 100 & 5000 & 200 \cr
5.96 & $24^3\times 48$ & 0.589159(3)    & $0.1000(5)$  & 2.40   & 5.002(25) & 100 & 2000 & 200 \cr
6.17 & $32^3\times 64$ & 0.610867(1)    & $0.0708(3)$  & 2.27   & 7.061(35) & 100 & 2000 & 200\cr
6.42 & $48^3\times 96$ & 0.632217(1)    & $0.0500(2)$  & 2.40   & 10.00(5)  & 100  & 2000 & 200\cr
\hline
\end{tabular}
\end{table*}
\subsection{Lattice setup}
We perform the pure Yang-Mills lattice simulations using the
standard Wilson plaquette action with a fixed physical volume
($La\approx 2.4$ fm) at four different gauge couplings ($\beta=6/g_0^2=
5.76$, 5.96, 6.17, and 6.42). 
Three of four ensembles ($\beta=5.96$, 6.17, and 6.42) (which correspond 
to the same lattice setups as in the original work of the Wilson 
flow done by L\"uscher~\cite{Luscher:2010iy}) had been generated 
for our previous study of tree-level improved lattice gradient flow~\cite{Kamata:2016any}.
In this study, we additionally generate a coarse lattice ensemble at $\beta=5.76$. 
The gauge configurations in each simulation are separated by
$n_{\rm update}$ sweeps after $n_{\rm therm}$ sweeps for thermalization
as summarized in Table~\ref{tab:Sims}.
Each sweeps consists of one heat bath~\cite{Cabibbo:1982zn} combined with four over-relaxation~\cite{Creutz:1987xi} steps. The number of configurations analyzed is 100
in each ensemble. All lattice spacings are set by the Sommer scale ($r_0=0.5$ fm).

%
%
\begin{figure*}[ht]
\centering
\includegraphics*[width=.48\textwidth]{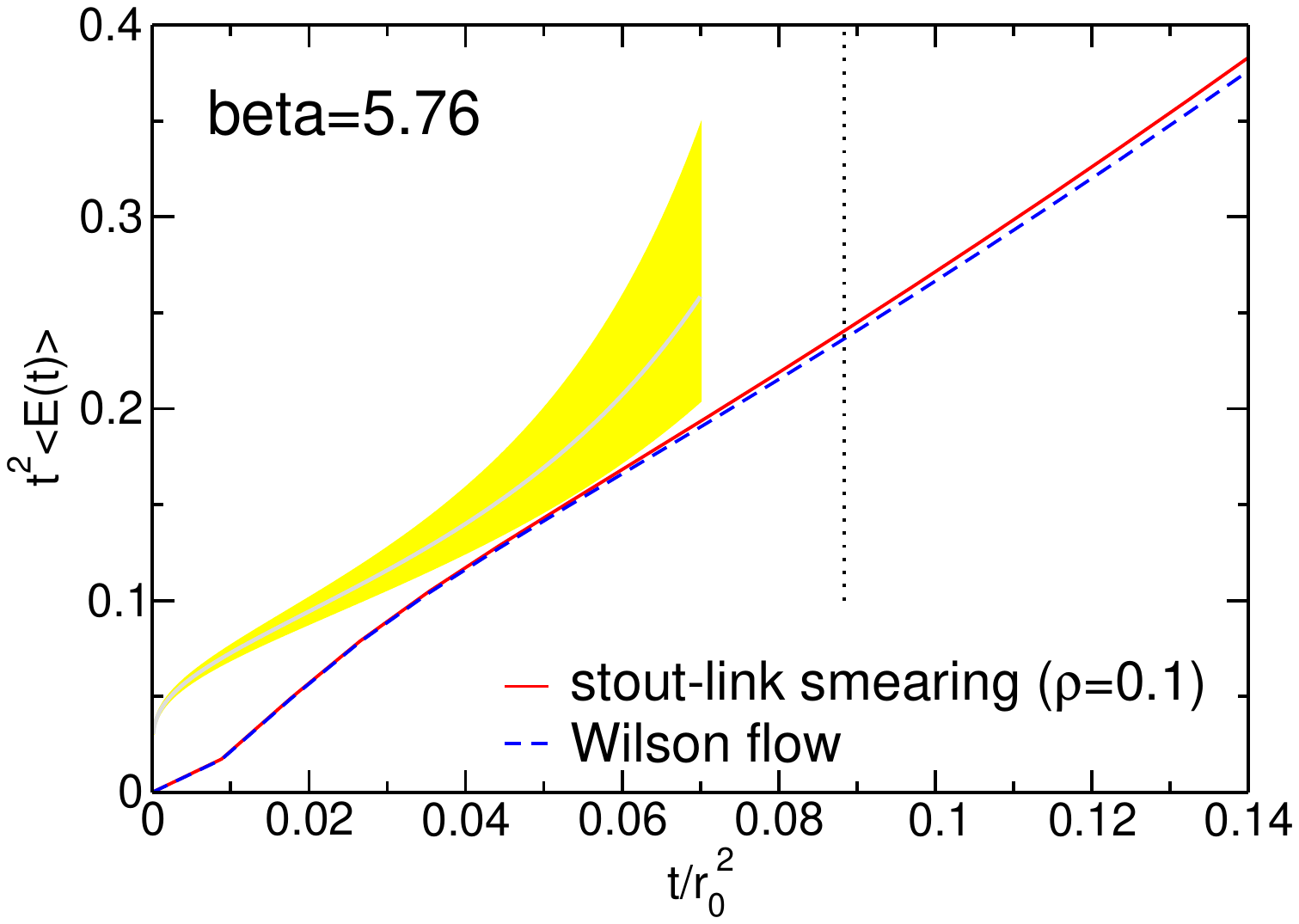} 
\includegraphics*[width=.48\textwidth]{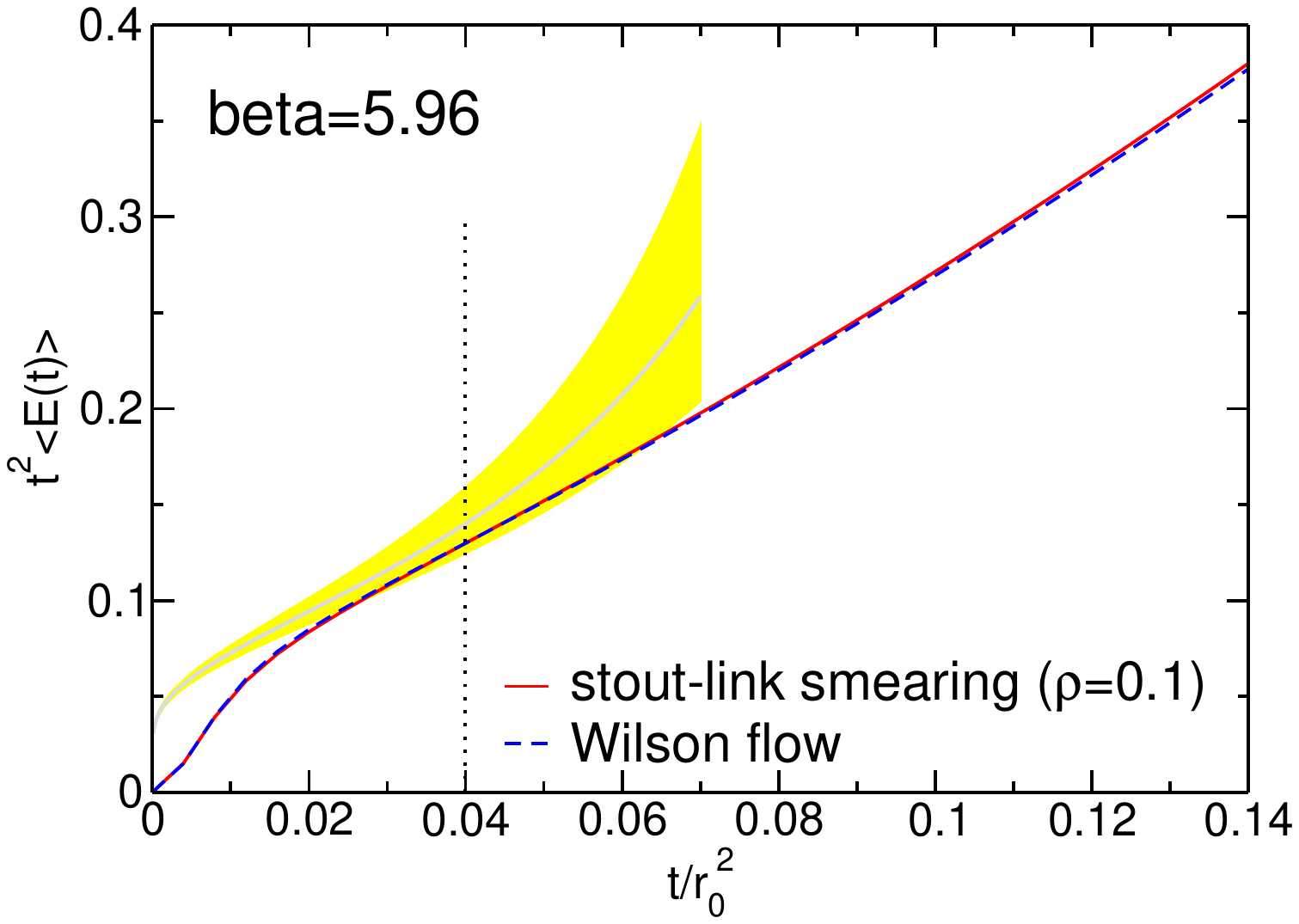} 
\includegraphics*[width=.48\textwidth]{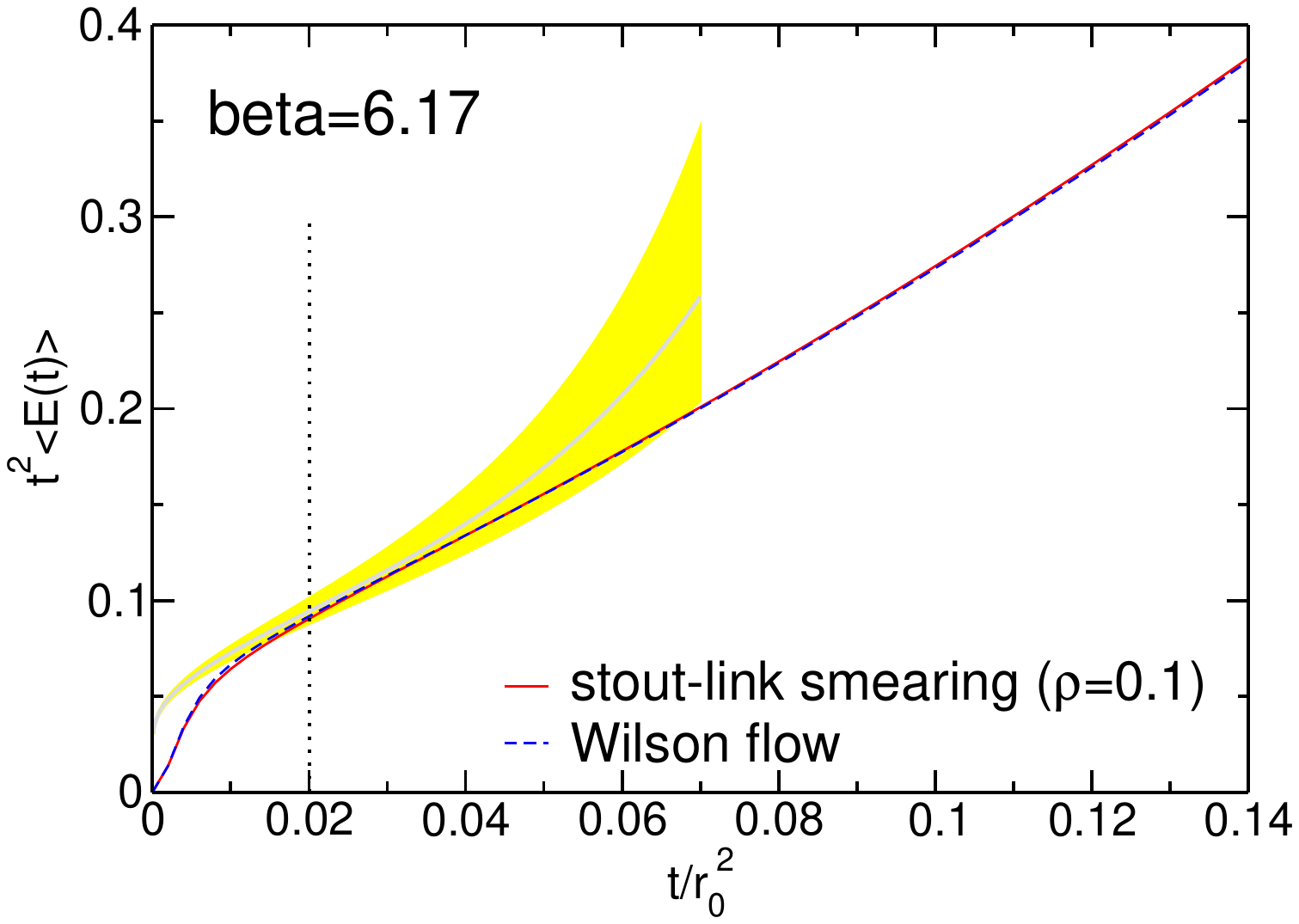} 
\includegraphics*[width=.48\textwidth]{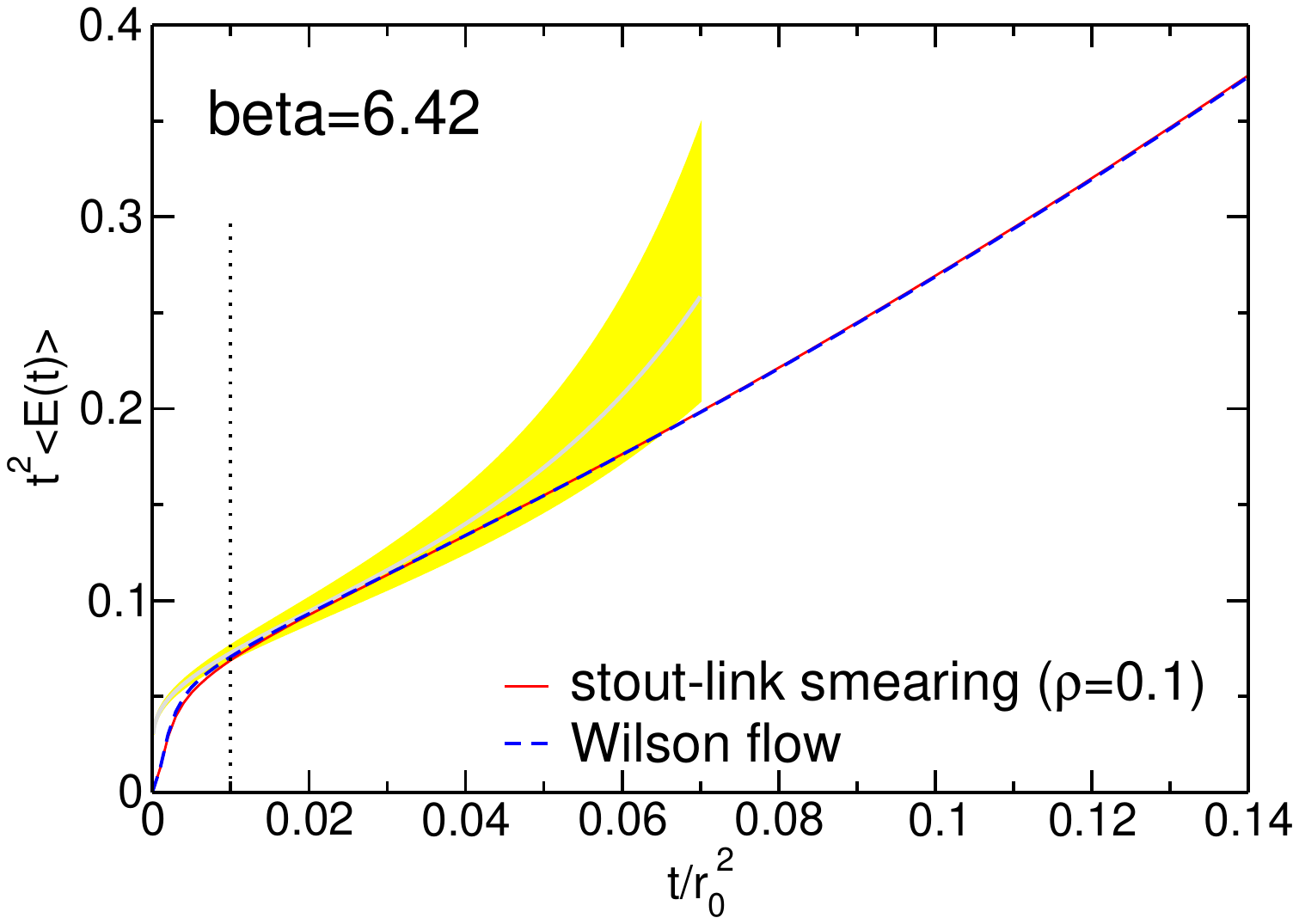} 
  \caption{The behavior of $t^2\langle E(t)\rangle$ computed by the Wilson flow (blue dotted curve) and the stout-link smearing with $\rho=0.1$ (red solid curve) at $\beta=5.76$ (upper-left panel),  $\beta=5.96$ (upper-right panel), $\beta=6.17$ (lower-left panel), and $\beta=6.42$ (lower-right panel).
  The gray solid curve with the yellow band corresponds to the continuum perturbative calculation~\cite{Luscher:2010iy} in each panel.}
\label{fig:WG}
\end{figure*}

\subsection{Direct comparison between the stout-link smearing and the Wilson flow}

We determine $ t^2 \langle E(t) \rangle$ 
using the expectation value of the clover-type action density $E(t, x) =\frac{1}{2}
{\rm Tr}\{G^{\rm cl}_{\mu\nu}(t, x)G^{\rm cl}_{\mu\nu}(t, x)\}$ with the clover-leaf operator $G^{\rm cl}_{\mu\nu}(t, x)$~\cite{Luscher:2010iy}.
The Wilson flow scale $t_0$ is given by the solution of the following equation~\cite{Luscher:2010iy}
%
%
\begin{equation}
\left. t^2 \langle E(t) \rangle \right|_{t=t_0}= 0.3.
\label{eq:new_scale}
\end{equation}

In this study, the forth-order Runge-Kutta scheme is used for the Wilson flow where the flow time $t$ is
given as $\epsilon \times n_{\rm flow}$ using the number of flow iterations $n_{\rm flow}$. 
As for the Wilson flow we evaluate $t^2 \langle E(t) \rangle$, 
denoted as $X_{\rm flow}(t)$, using $\epsilon=0.025$ to compare with the stout smearing results after we check that $\epsilon=0.1$, 0.025, and 0.01 give rise consistent results. As for the stout smearing, we evaluate $t^2 \langle E(t) \rangle$, denoted as $X_{\rm stout}(t)$, with $\rho=0.1$, 0.025, and 0.01 for each ensemble
listed in Table~\ref{tab:Sims}, where the flow time $t$ corresponds to $\rho n_{\rm st}$.
Figure~\ref{fig:WG} shows typical behaviors of 
$X_{\rm flow}(t)$ (blue dotted curve) and 
$X_{\rm stout}(t)$ for the smearing parameter $\rho=0.1$ (red solid curve).

\begin{figure*}
\centering
\includegraphics*[width=.48\textwidth]{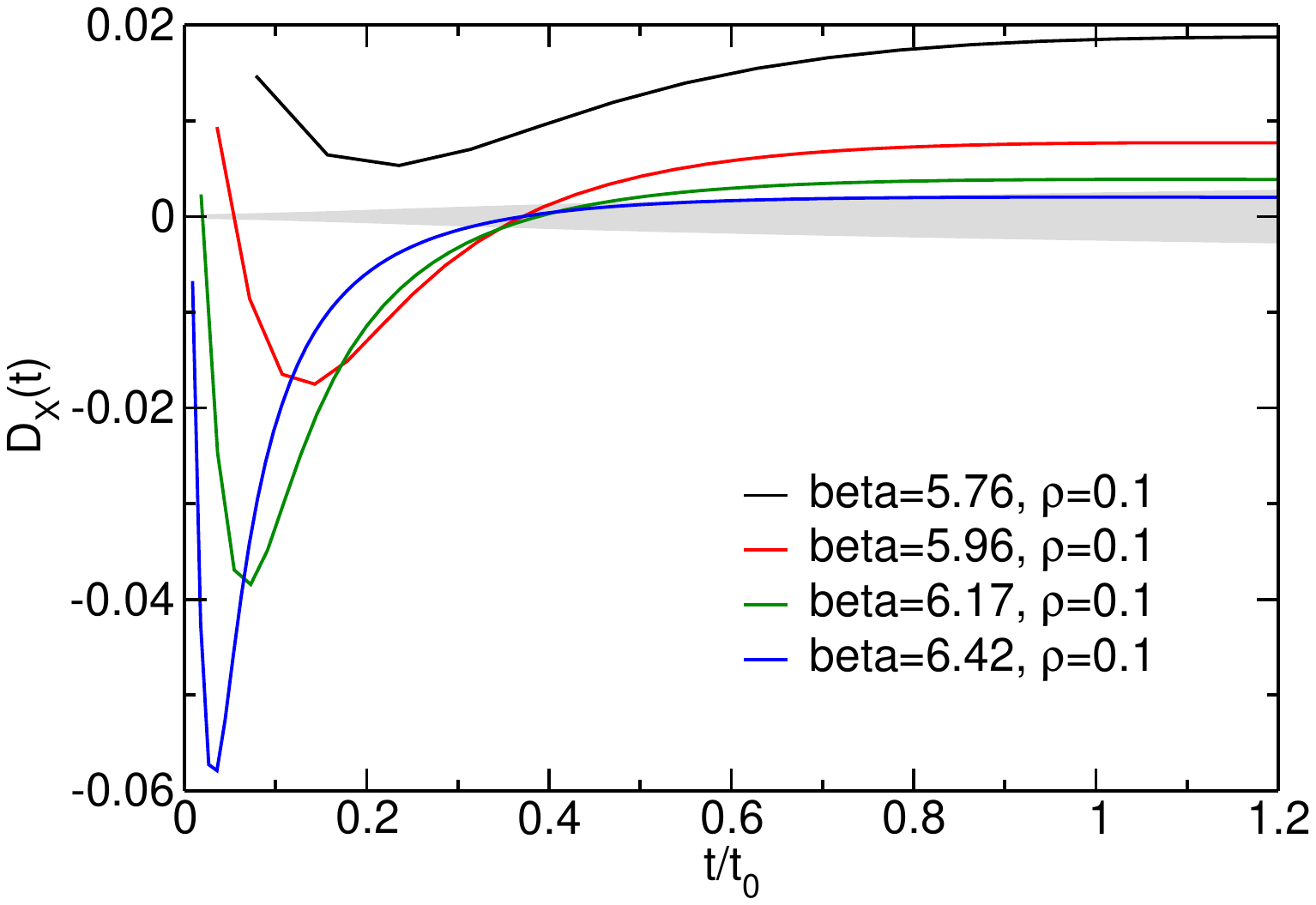} 
\includegraphics*[width=.48\textwidth]{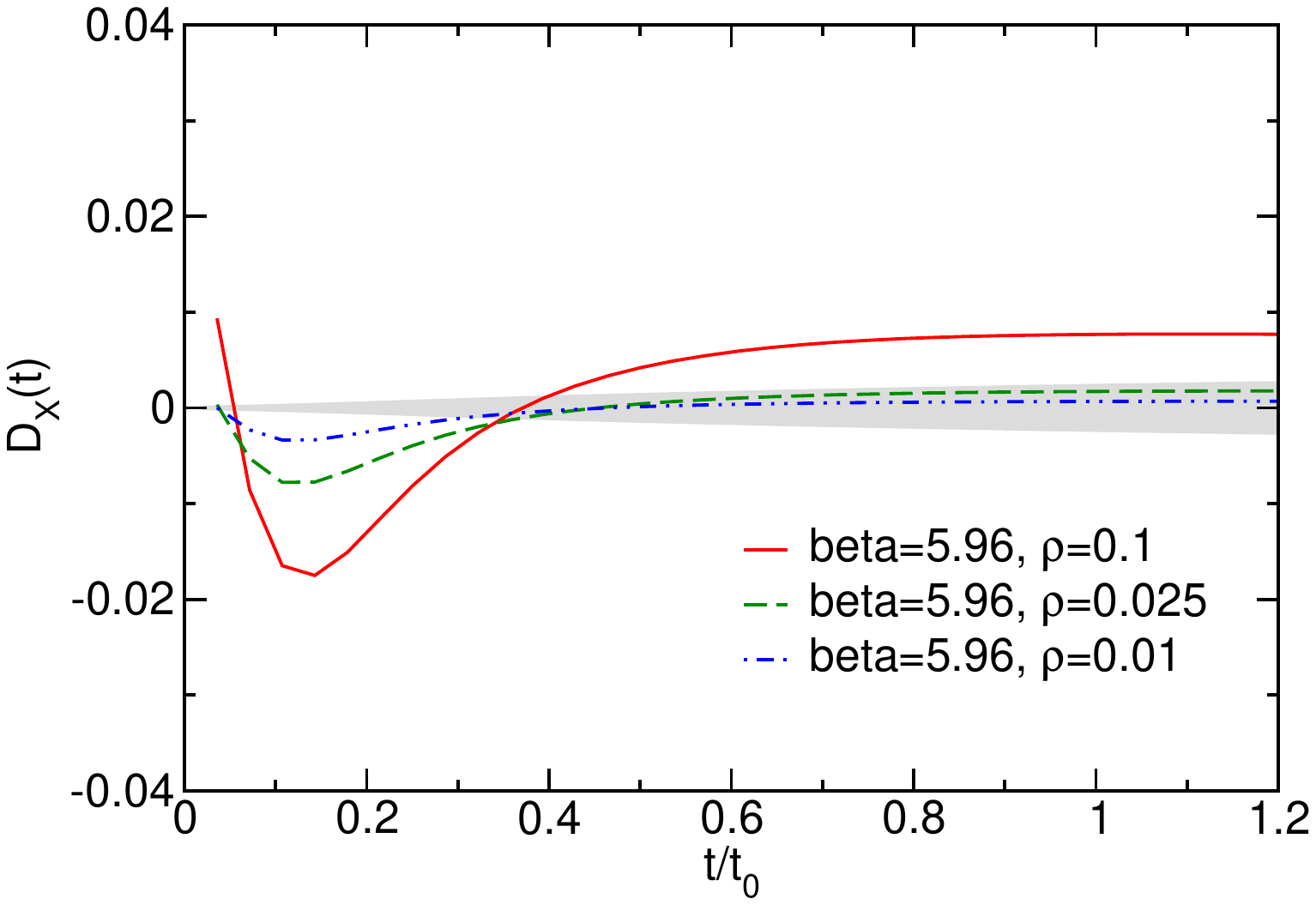} 
\caption{
{\bf Left}: The lattice spacing dependence of $D_X(t)$ calculated between the Wilson flow and the stout smearing ($\rho=0.1$) at $\beta=5.76$, 5.96, 6.17, and 6.42.
The area shaded in gray corresponds to the relative size of the statistical uncertainties
on $X(t)$ determined by the Wilson flow in this study.
{\bf Right}:
The behavior of 
$D_X(t)$ obtained with three smearing parameters:
$\rho=0.1$ (solid curve), $\rho=0.025$ (dashed curve),
and $\rho=0.01$ (double-dotted curve) 
as functions of $t/t_0$ calculated at $\beta=5.96$.
The area shaded in gray corresponds to the relative size of the statistical uncertainties on $X(t)$ obtained by the Wilson flow.} 
\label{fig:X_b596}
\end{figure*}
Remark that the stout smearing almost reproduces the behavior of $X_{\rm flow}(t)$ even when the Wilson flow certeinly deviates from the perturbative calculation of the continuum Yang-Mills gradient flow.

The detailed discussion can be made by defining the following ratio:
%
%
\begin{equation}
D_X(t)\equiv\frac{X_{\rm stout}(t)-X_{\rm flow}(t)}{X_{\rm flow}(t)}.
\end{equation}

The left panel of Fig.~\ref{fig:X_b596} shows the behavior of $D_X(t)$ for all four $\beta$ values with the smearing parameter $\rho=0.1$ and the right panel of Fig.~\ref{fig:X_b596} shows the behavior of $D_X(t)$ for three smearing parameters at $\beta=5.96$ as a function of $t/t_0$.
Although peak structures appear at the small-$t$ region ($t/t_0< a^2/t_0$), all $D_X(t)$ saturate at the large-$t$ region ($t/t_0 > a^2/t_0$).

\begin{figure*}[ht]
\centering
\includegraphics*[width=.8\textwidth]{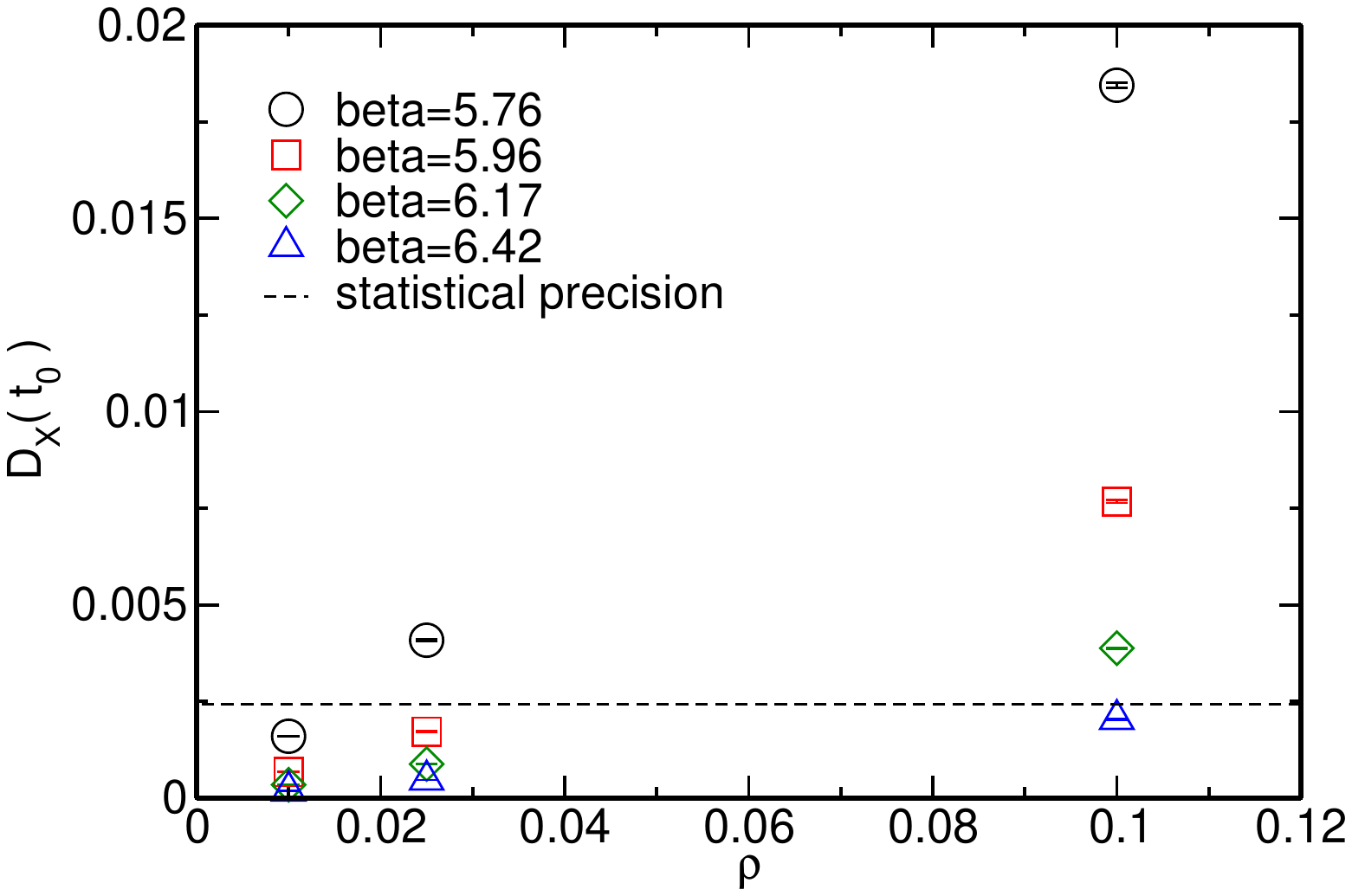} 
\caption{ 
The smearing parameter dependence of $D_X(t)$ evaluated at $t=t_{0}$
with three smearing parameters: $\beta=5.76$ (circles) ,
$\beta=5.96$ (squares), $\beta=6.17$ (diamonds), $\beta=6.42$ (upper triangles). 
A horizontal dotted line represents the relative size of the statistical uncertainties ($\sim 0.24\%$) on
the values of $t_0$ calculated by the Wilson flow in this study.}

\label{fig:diff_smear}
\end{figure*}

Figure~\ref{fig:diff_smear} shows $D_X(t_0)$ for any combinations of parameters to evaluate the parameter dependence of the saturated values. It is clear from Fig.~\ref{fig:diff_smear} that $D_X(t)$ tends to be smaller as the lattice spacing and the stout smearing parameter decrease as expected from what we learned in Sec.~\ref{sec:FLOW_STOUT}. The horizontal dotted line in Fig.~\ref{fig:diff_smear} represents the statistical uncertainties divided by observed values with respect to $X_{\rm flow}(t)$.

From these observations, we claim that we can numerically identify the stout smearing as the Wilson flow if we choose the proper combination of the parameters. Especially, $\rho=0.1$ for $\beta=6.42$, $\rho=0.025$ for $\beta=5.96$ and 6.17, and $\rho=0.01$ for $\beta=5.76$ are the enough small stout smearing parameter.

%
%

\section{Summary} 
\label{sec:SUMMARY}
In this talk, we presented the prospects on the stout smearing as an equivalent approach to the Wilson flow.
First, we show the proof of the analytical equivalence at finite lattice $a$ in the zero limit of the stout-smearing parameter $\rho$. The proof also implies that finite $\rho$ induces ${\cal O}(a^2)$ correction to the equivalence. Second, to verify whether they remain equivalent in numerical simulations within the statistical precision, we have performed the numerical simulations with the Wilson gauge configurations generated at four different gauge couplings ($\beta=5.76$, 5.96, 6.17, and 6.42). The direct comparison of the expectation values of the action density indicates that $\rho=0.1$ for $\beta=6.42$, $\rho=0.025$ for $\beta=5.96$ and 6.17, and $\rho=0.01$ for $\beta=5.76$ are enough small to identify the stout smearing as the Wilson flow in the determination of the reference scale $t _0$.

As for the prospects on the equivalence, we suggest two applications. One is the use of the stout smearing as an alternative method of the Wilson flow in the numerical simulations, which could reduce a factor of ${\cal O}(10)$ computational costs for the same iteration steps. We also consider that the sophisticated perturbation analysis in the gradient formalism can be used to calculate the one loop quantities in lattice perturbation theory for the smeared-link fermion action.

\begin{acknowledgments}
We would like to thank M. Ammer and R.J. Hudspith for useful discussions.
M. N. and K. S. are supported by Graduate Program on Physics for the Universe (GP-PU)
of Tohoku University. 
Numerical calculations in this work were partially performed using Yukawa-21 
at the Yukawa Institute Computer Facility,
and also using Cygnus at Center for Computational Sciences (CCS), University of Tsukuba
under Multidisciplinary Cooperative Research Program of CCS (MCRP-2021-54, MCRP-2022-42). 
This work was also supported in part by Grants-in-Aid for Scientific Research form the Ministry 
of Education, Culture, Sports, Science and Technology (No. 18K03605 and No. 22K03612).
\end{acknowledgments}

\end{document}